\documentclass[twocolumn,showpacs,prl]{revtex4}
\usepackage{graphicx}

\begin{document}
\title{Quantum Hall Ferrimagnetism in lateral quantum dot molecules}

\author{Ramin M. Abolfath}

\author{Pawel Hawrylak}

\affiliation{Institute for Microstructural Sciences,
National Research Council of Canada,
Ottawa, K1A 0R6, Canada}

\date{\today}

\begin{abstract}
We demonstrate the existance of ferrimagnetic and ferromagnetic phases in a spin phase diagram of  
coupled lateral quantum dot molecules in the quantum Hall regime.  
The spin phase diagram is determined from Hartree-Fock Configuration Interaction method
as a function of electron numbers N, 
magnetic field B, Zeeman energy, and tunneling barrier height. The quantum Hall ferrimagnetic phase
corresponds to spatially 
imbalanced spin droplets resulting from strong inter-dot coupling of identical dots. The
quantum Hall ferromagnetic 
phases correspond to ferromagnetic coupling of spin polarization at filling factors 
between $\nu=2$ and $\nu=1$. 
\end{abstract}

\maketitle


Electron-electron interactions are responsible for the magnetic properties
of solids\cite{mattis,Ferri}. It is now possible to engineer the many-body states of electrons 
by confining them in artificially fabricated nanostructures to control their 
magnetic 
properties.\cite{OhnoDietlNature,sachrajda hawrylak ciorga,Mn qdots}
This has been demonstrated in quantum wells\cite{QHF} and dots where by varying the magnetic field
the total electron spin has been tuned from $S=0$ at a spin singlet 
quantum Hall droplet at filling factor 
$\nu=2$ to maximally spin polarized ferromagnetic 
 filling factor $\nu=1$ droplet.\cite{tarucha,Ciorga,Hanson03}
More complex spin textures associated with correlated states such as spin 
bi-excitons were also identified and observed experimentally \cite{Korkusinski}.

Given the tunability of total spin with magnetic field and electron numbers 
in a single dot it is natural to explore
the possibility of magnetic coupling in quantum dot molecules.  
\cite{palacios,kouwenhoven,petta,RM,MichelPRL93,Hatano,RaminWojtekPawel}
In this Letter we explore the effect  of  magnetic field, interdot tunneling
and electron-electron interactions on the evolution of total electron spin for
different electron numbers in a lateral quantum dot
molecule. The inter-dot and intra-dot 
electron-electron Coulomb interactions 
are incorporated systematically using the Hartree-Fock Configuration 
Interaction method (HF-CI). 
We find quantum Hall droplets  with zero and full spin polarization, 
identified as $\nu=2$ and
$\nu=1$ quantum Hall droplets 
in analogy with single quantum dots
and quantum Hall ferromagnetism \cite{QHF}.
Between these two states, we find 
series of continuous transitions among partially spin polarized phases.
Simultaneous 
spin flips in each isolated dot must lead to even number of spin flips 
in a double dot. 
Surprisingly, we find partially polarized phases
with odd number of spin flips. 
We identify these correlated states as quantum Hall {\em ferrimagnets},
\cite{Ferri} which are a direct manifestation of coherent
quantum mechanical coupling between the quantum dots as well as inter-dot
electronic correlations.


\begin{figure}
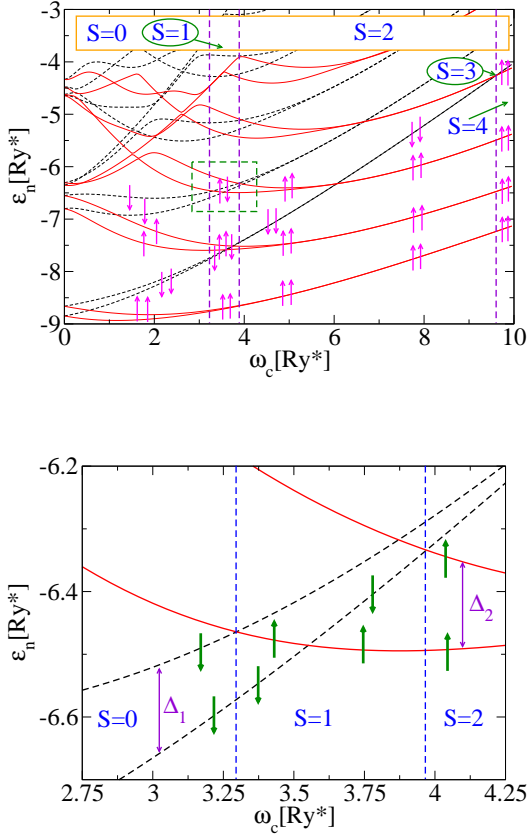

\begin{center}
\includegraphics[width=0.8\linewidth]{E_g0.62_1_0.eps}\vspace{1cm}
\includegraphics[width=0.8\linewidth]{E_g0.62_1_1.eps}
\noindent
\caption{Double dot single particle spectrum in the presence of 
Zeeman coupling. 
For illustration purposes a very high Zeeman coupling $g=-9$
was used.
The inset shows the evolution of total spin as a function of 
magnetic field in a quantum dot molecule with $N=8$ 
non-interacting electrons.
The states surrounded by circles show odd spin.
The stability range of these states is very narrow due to small inter-dot
tunneling amplitude. 
The bottom is the enlarged part of the spectrum 
(the box in the top) which shows 
the evolution of first and second spin flip transitions.
}
\label{spectrum}
\end{center}
\end{figure}

We describe electrons confined in quasi-two-dimensional 
quantum dot molecule in a uniform perpendicular 
magnetic field by the effective mass Hamiltonian
$H = \sum_{i=1}^N  T_i
+ \frac{e^2}{2\epsilon}\sum_{i \neq j}\frac{1}{|\vec{r}_i - \vec{r}_j|} +
E_Z$,  
where
$T=\frac{1}{2m^*}\left(\frac{\hbar}{i}\vec{\nabla}
+ \frac{e}{c} A(\vec{r})\right)^2 + V(x,y)$
is the single electron Hamiltonian in magnetic field. Here $(\vec{r})=(x,y)$
describes electron position, $A(\vec{r})=\frac{1}{2}\vec{B}\times\vec{r}$ 
is the magnetic vector potential, and
$m^*$ is the conduction-electron effective mass, 
$e$ is the electron charge, and
$\epsilon$ is the host semiconductor dielectric constant 
($\epsilon=12.8$ in GaAs).
$E_Z=\gamma \omega_{c}$ is the Zeeman energy, $\sigma= +1~(-1)$ 
corresponds to spin $\uparrow(\downarrow)$, $\gamma=m^* g$, 
$g$ is the host semiconductor $g$-factor ($g=-0.44$ in GaAs and $g=-14$ 
in InAs), 
and $\mu_B$ is the Bohr magneton.
$V(\vec{r})$ is the quantum dot molecule confining potential,
parameterized in terms of a sum of three Gaussians   
$V(x,y)=V_0~ \exp[{-\frac{(x+a)^2+y^2}{\Delta^2}}]
        +V_0~ \exp[{-\frac{(x-a)^2+y^2}{\Delta^2}}]
+V_p \exp[{-\frac{x^2}{\Delta_{Px}^2}-\frac{y^2}{\Delta_{Py}^2}}]$.
Here $V_p$ is the 
central plunger gate potential controlling the tunneling barrier.
The parameters $V_0=-10, \Delta=2.5, a=2$, 
and $\Delta_{Px}=0.3$, $\Delta_{Py}=2.5$ in effective atomic units 
describe a typical confining potential 
corresponding to weakly coupled quantum dots.
Diagonalizing single particle Hamiltonian as in Ref.\onlinecite{RaminWojtekPawel}
we obtain single particle spectrum as a function of magnetic field 
  shown in Fig. \ref{spectrum}. In order to illustrate the physics of spin flips
in a coupled molecule the spectrum corresponds to a
large Zeeman coupling.
At zero magnetic field it exhibits well separated S,P, and D electronic shells,
and at high magnetic fields it forms lowest Landau level molecular
shells of closely spaced pairs of bonding-antibonding orbitals. 
In high magnetic field the corresponding wavefunctions admit a 
description in terms of localized LLL orbitals \cite{RaminWojtekPawel}. 
In this limit linear combinations of the LLL orbitals $m$ 
from left and right dot forms molecular
shells of closely spaced symmetric-antisymmetric pairs with
eigen-energies expressed approximately as
$\epsilon_{m\lambda\sigma} = 
\omega_{-}(m+\frac{1}{2}) 
- \lambda \frac{\Delta_m}{2} - \frac{1}{2}\sigma \gamma \omega_{c}$ where
$\omega_{-}=\sqrt{\omega_0^2 + \omega_c^2/4} - \omega_c/2$
(Fock-Darwin eigen-energies of a parabolic dot), and
$\omega_0=2\sqrt{|V_0|/\Delta^2}$ \cite{RaminWojtekPawel}.
Here $\lambda$ is the pseudospin index,
the symmetric (antisymmetric) orbitals labeled by $\lambda=+1~(-1)$.
$\Delta_m$ is the symmetric-antisymmetric gap.
These orbitals can be
grouped into electronic shells, depending on the number of electrons.
The half-filled molecular shells correspond to electron numbers 
$(N_L=2k-1,N_R=2k-1)$ and filled shells correspond to 
$(N_L=2k,N_R=2k)$ configurations ($k$ is integer).
 

In Fig. \ref{spectrum} $S_z=0,2$, and $4$ phases of the $N=8$ electron droplet
are shown. They are obtained by occupying the lowest single particle energy levels.
The even $S_z$ phases can be understood 
as simultaneous spin flips in each isolated dot. 
However, we see in Fig.\ref{spectrum} also odd $S_z$ phases. 
In the bottom part of Fig. \ref{spectrum} the first odd spin $S_z=1$ state
occurs between magnetic fields corresponding to 
$\omega_c(1) \approx 3.25$ and $\omega_c(2) \approx 3.9$, which in turn
corresponds to the crossing of energy levels 
$\epsilon_{m=2,\sigma=\uparrow,\lambda=+1}=
 \epsilon_{m=1,\sigma=\downarrow,\lambda=-1}$, and
$\epsilon_{m=2,\sigma=\uparrow,\lambda=-1}=
 \epsilon_{m=1,\sigma=\downarrow,\lambda=+1}$.
Using single particle eigen-energies, we find the stability
range of the odd spin phase  directly related to 
the tunneling splitting of energy levels:
$\gamma[\omega_c (2) -  \omega_c (1)] = [\Delta_2(2)+\Delta_1(2)]/2 
+ [\Delta_2(1)+\Delta_1(1)]/2$.
Hence the odd spin flip state is related to the 
splitting of energy levels due to tunneling:
States with $S_z=1$, and $S_z=3$,
are stable within narrow range of magnetic fields  
due to spin flip transitions among the electrons
that occupy the levels with energy separation proportional to
the inter-dot tunneling amplitude.
This is in contrast with the first spin flip transition in single dots
which is stable in a wide range of magnetic fields.
For this reason the existence and stability range of odd spin states
in the spin phase diagram of 
quantum dot molecules can be interpreted as the measure of 
inter-dot interaction.

\begin{figure}
\begin{center}
\includegraphics[width=0.8\linewidth]{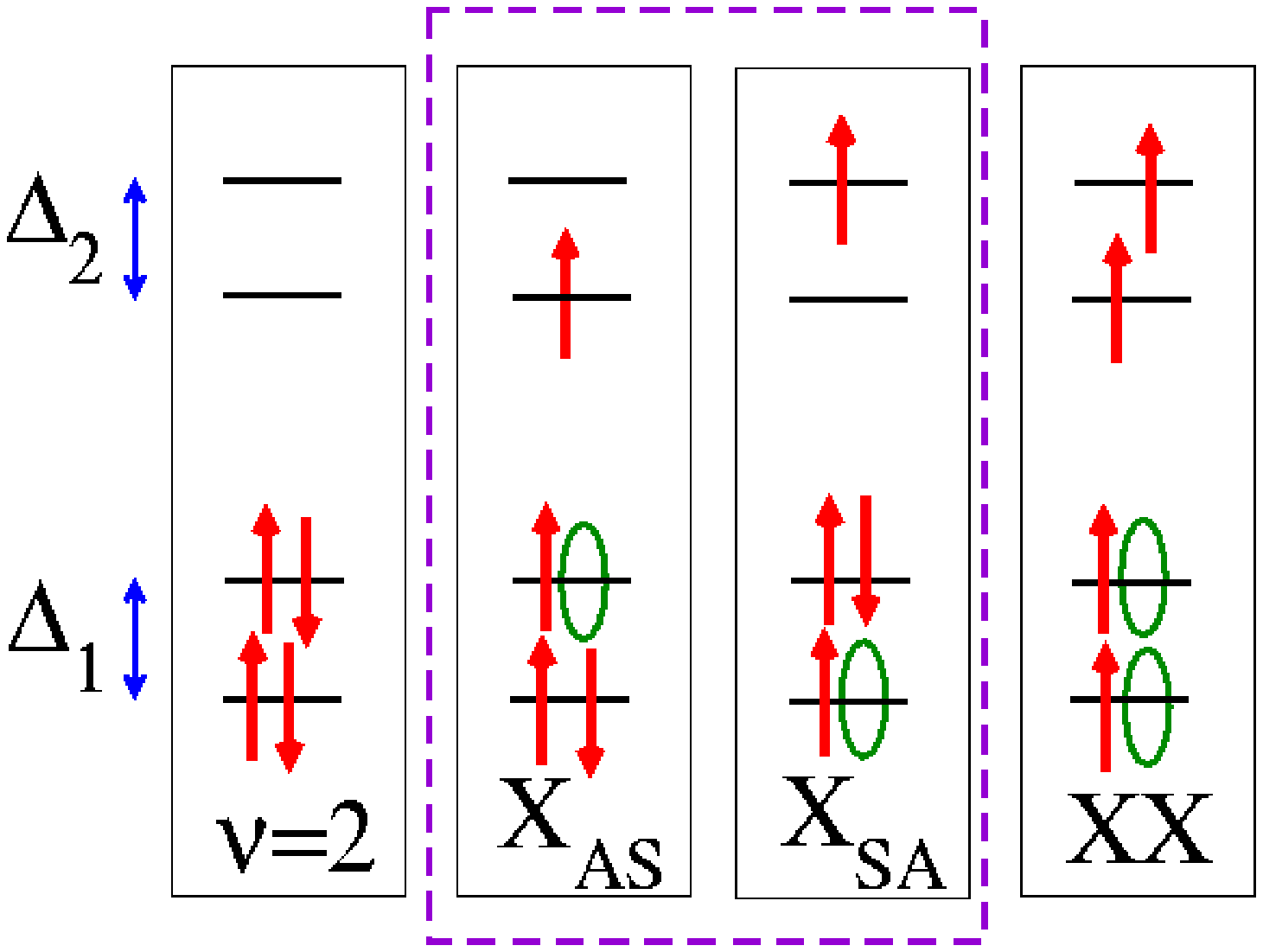}\vspace{1cm}
\includegraphics[width=0.8\linewidth]{T5_1_forPRL.eps}\vspace{1cm}
\noindent
\caption{
(Top) The basis of spin configurations in high magnetic fields.
The first spin transition states $S=1$ identify with two independent
set: $\{X_{SS},X_{AA}\}$, and $\{X_{SA},X_{AS}\}$.
In the former the electron-hole transitions occurs between the
states with the same symmetry and hence
they do not mix with the latter which exhibit the process of 
electron-hole excitations between states with opposite parity. 
(Bottom) The energies of two  single spin excitons $X_{SA}$ and $X_{AS}$ 
with odd parity, the energy of the odd parity correlated exciton 
$X_{SA+AS}$, and the energy of the spin bi-exciton $S=2$ state as a 
function of magnetic field. All energies measured from the energy of the 
$\nu=2,S=0$ state.
}
\label{Xsa}
\end{center}
\end{figure}


The above analysis emphasizes the relationship between odd spin phases and 
quantum mechanical coupling between the dots. This coupling in single dots
is not due to single particle effects but due to electron-electron interaction.
Hence we now turn off Zeeman coupling and analyse the effects of electron-electron 
interactions on spin transitions. 
We focus on the tunnel coupled lowest Landau level orbitals $m$.
The $\nu=2$  state of quantum dot molecule with $N$
electrons and total spin $S=0$ is shown in Fig. \ref{Xsa}.
The energy associated with this state 
$E_{\nu=2}= \sum_{m=0}^{N/4 -1} \sum_{\lambda =\pm 1} \sum_{\sigma}
\left[\epsilon_{m\lambda\sigma} + \Sigma(m,\lambda,\sigma)\right]$
can be expressed in terms of electron self-energy $\Sigma(m,\lambda,\sigma)$. 
The $S=1$ spin flip excitation is constructed by removing an electron
from occupied $\nu=2$ state and putting into an unoccupied state.
There are two possible spin excitons for a given parity,
depicted in Fig. \ref{Xsa}. 
Fig. \ref{Xsa}(bottom) shows the numerically calculated 
energies of spin excitons in  $N=8$ electron droplet as a function of magnetic field.
The energy $\Delta E_{X_{AS}}$  of spin exciton  $X_{AS}$ is positive 
for magnetic fields shown
but the energy $\Delta E_{X_{SA}}$ of spin exciton $X_{SA}$ becomes negative 
at $\omega_c=3.8$ i.e.
the  $X_{SA}$ spin flip state becomes the lower energy state than the 
$\nu=2, S=0$  state. 
However, in stark contrast with a single quantum dot 
we find that the second spin flip state $XX$ becomes 
the ground state
at lower magnetic field $\omega_c=2.1$. Hence, unlike in a single quantum dot
we find a transition from spin singlet $S=0$ state directly to $S=2$ second 
spin flip state.
This is a transition corresponding to even total spin numbers, as if the two 
dots were flipping their
spin simultaneously. However, the two spin excitons 
$X_{AS}$ and
$X_{SA}$ are not eigenstates of the Hamiltonian and are coupled
by Coulomb interaction. The resulting energy 
$\Delta E_{X_{SA}+X_{AS}}$ of the correlated 
single spin state is significantly lower and equals both the energy 
$\Delta E_{XX}$ of the second spin flip bi-exciton and of  the $\nu=2, S=0$  
state at $\omega_c=2.1$. 
At this magnetic field
the energy of the bi-exciton and of the exciton are almost identical. 
The $S=1$ states show stability in a narrow range of magnetic
field, within the accuracy of our numerical results.
With further increase of magnetic field, 
single excitons condense into pairs of excitons forming biexcitons.
The existence of single, odd, spin excitons is hence a signature of 
electronic correlations. 
These states can be thought of as one of the quantum dots underwent 
spin transition and the excess spin tunneled back and forth the two dots.
The correlated odd spin state does not exist at the Hartree-Fock level as shown above
and   
can be identified as quantum Hall ferrimagnet.

\begin{figure}
\begin{center}
\includegraphics[width=0.8\linewidth]{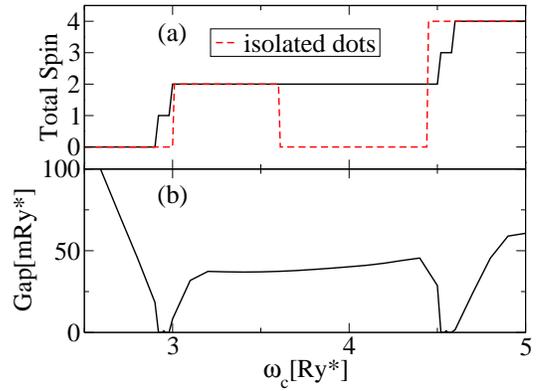}\vspace{1cm}
\noindent
\caption{
(a) The evolution of spin  of the $N=8$ electron 
quantum dot molecules as a function of magnetic field for 
$\gamma=-0.02$ and $V_p=7$ 
(solid line). 
For comparison the spin evolution of two isolated dots with zero Zeeman
splitting is shown (dashed line).
(b) The evolution of the energy gap  of the 
quantum dot molecules as a function of magnetic field.
} 
\label{Gap1}
\end{center}
\end{figure}

To support the assertion that correlations are responsible
for the existence of odd spin excitons we employ Hartree-Fock (HF)
configuration interaction (CI) method to calculate the ground state properties
of quantum dot molecules in a magnetic field and with electron numbers up to  $N=12$.
The HF basis is used for the construction of multi-electron configurations, which in
turn are used for the construction of the Hamiltonian matrix.
The Hamiltonian matrix is either diagonalized exactly for small systems,
or low energy eigenvalues and eigenstates are extracted approximately 
for very large number of configurations ($\approx 10^7$) \cite{RaminPawel}.
Fig. \ref{Gap1} shows the evolution of total spin and the energy gap
with increasing magnetic field in $N=8$ quantum dot molecule.
For comparison the spin evolution of two noninteracting $N=4$ quantum 
dots is shown with a dashed line. 
We find that the effect of 
direct, exchange, and correlation energies calculated by HF-CI
is to renormalize the 
magnetic fields at which spin transitions take place, and more importantly,
to lead to the appearance of odd spin states $S=1$ and $S=3$, in agreement with
our previous analysis.

While the existence of odd spin states
is most striking, the presence of spin polarized phases is also nontrivial. 
The fact that
spins of electrons on two quantum dots align demonstrates the existence of 
ferromagnetic 
dot-dot coupling. In the case of antiferromagnetic coupling there would have 
been no
net spin eventhough each dot has finite spin. 
As shown in Fig. \ref{spectrum} without electron-electron interactions 
competition between 
quantum mechanical tunneling and Zeeman energy was responsible for the 
existence
of odd spin phases. The effect of finite Zeeman energy is similar in an 
interacting system.
The effect of increasing Zeeman energy on the evolution of spin  
and  energy gap of the $N$-electron quantum dot molecules as a function 
of magnetic field is to 
renormalize the magnetic field value of spin flips and, more importantly,
to stabilize the odd spin phases.

\begin{figure}
\begin{center}
\includegraphics[width=0.9\linewidth]{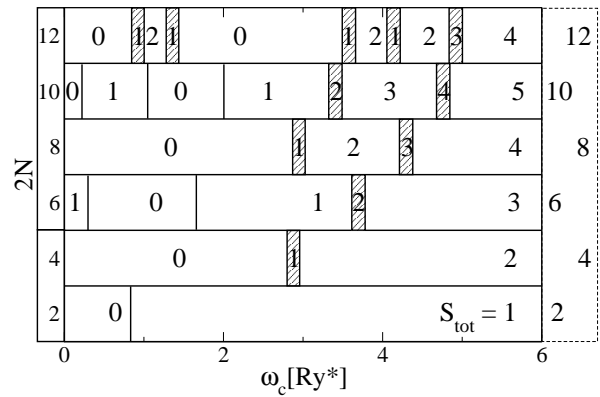}
\noindent
\caption{Spin phase diagram of artificially coupled dots
molecules in  the presence of Zeeman coupling with $\gamma=-0.02$ and 
tunneling barrier $V_p=7$.
Numbers in blocks represent the total spin.
Shaded areas show the stability region of ferrimagnetic states.
}
\label{URHFCI_phase_diagram2}
\end{center}
\end{figure}

Here we summarize our results for other even electron numbers, 
$N=2,4,6,8,10$, and $12$, in the form of a phase diagram shown in 
Fig. \ref{URHFCI_phase_diagram2}.
The tunneling barrier and Zeeman coupling are $V_p=7$ and $\gamma=-0.02$.
With the choice of electron numbers the phase diagram covers quantum molecules
build out of artificial atoms with up to second filled shell. For example,
$N=4$ molecule corresponds to two quantum dots $(2,2)$, each with a filled s-shell.
The $N=12$ molecule is build of two quantum dots $(6,6)$, each with a filled p- and 
s-shell. Intermediate electron numbers correspond to molecules build out of 
artificial atoms with partially filled shells. Partial shell filling implies
Hunds rules and total electron spin, e.g. $S=1$ for a four electron quantum dot.
When the two dots $(4,4)$ with four electrons each are brought together into
$N=8$ electron molecule, the total electron spin of the ground state is 
found to be $S=0$,
as seen in Fig. \ref{URHFCI_phase_diagram2}. But when the two dots $(3,3)$ with three
 electrons each are brought together into
$N=6$ electron molecule, the total electron spin of the ground state is $S=1$.
Hence at zero magnetic field there exist a rich phase diagram sensitive to the
separation of the two dots and the barrier height. The situation is significantly
simplified by the application of the magnetic field.
For all electron numbers increasing magnetic field forces electrons 
to flip the spin and increase spin polarization until  all electron spins
are aligned. In the phase diagram the spin polarized phase with $S=N/2$  
is visible.
This phase can be attributed to the 
ferromagnetic $\nu=1$ quantum Hall droplets.
The spin transitions leading to the spin polarized droplet do depend on the number
of electrons. 
We discuss below the nature of the magnetic field induced spin transitions 
for half-filled shells of coupled dots with $(N_L+1,N_R+1)$ electron numbers.
The half-filled shells correspond to filled shells and additional electron
in each dot. For $(1,1)$ there are only $N=2$ electrons and one finds the singlet
at low magnetic fields followed by a triplet at higher magnetic fields.
For higher electron numbers  the system can be interpreted
as effective two-electron $(N_L=1,N_R=1)$ artificial molecules,  
with $S=0$ core electrons  $(S_L=0,S_R=0)$ and finite spin
valence electrons with antiferromagnetic/ferromagnetic coupling,
corresponding to spin singlet 
$|\uparrow\downarrow\rangle - |\downarrow\uparrow\rangle$,
and spin triplet
$|\uparrow\uparrow\rangle$ \cite{MichelPRL93,RaminWojtekPawel} states.
Thus the corresponding Hamiltonian can be reduced to 
the Heisenberg model $H=J S_L \cdot S_R$
by restricting the valence electrons to a
two level model (Hund-Mullikan approximation), 
and by  extension of the half-filled Hubbard model. 
Here $S_L$ and $S_R$ are spin-$1/2$ operator for the two localized
valence electrons, and $J$ is the singlet-triplet splitting.
The spin singlet $S=0$ state, visible in the phase diagram 
of Fig. \ref{URHFCI_phase_diagram2}
corresponds to filling factor $\nu=2$ quantum Hall droplets. 
Finally, in Fig. \ref{URHFCI_phase_diagram2}
small domains of quantum Hall ferrimagnetism 
with $S=$(even) and $S=$(odd) corresponding to 
half-filled and filled shells, stabilized by Zeeman energy, are shown.
The half-filled shells with $S$-even can be easily understood. For example,
for $N=10$ the $S=2$ phase corresponds to $S=1/2$ on one dot and $S=3/2$ on a second dot,
i.e. only the second dot underwent a spin flip.

In conclusion, 
we have presented the spin phase diagram of quantum dot
artificial molecules for different number of electrons as a function of external
magnetic field, Zeeman energy,
and tunneling barrier. The phase diagram was obtained using Hartree-Fock-Configuration
Interaction method. The magnetic field allows the tuning of the total spin of electrons in
each artificial atom. Quantum mechanical tunneling and electron-electron interactions
couple spins of each artificial atom and result in ferromagnetic and anti-ferromagnetic states
tunable by the magnetic field and barrier potential. 
Rather surprisingly,
ferri-magnetic states in which one of the dots undergoes spontaneous change of spin
were found and predicted.

Authors acknowledge the support by the NRC High 
Performance Computing project and by the Canadian 
Institute for Advanced Research.



\end{document}